\documentclass[10pt,conference]{IEEEtran}
\usepackage[T1]{fontenc}
\usepackage[utf8]{inputenc}
\usepackage{paralist}
\usepackage{booktabs}

\ifCLASSINFOpdf
\usepackage[pdftex]{graphicx}
\else
\fi
\usepackage[cmex10]{amsmath}
\usepackage{fixltx2e}
\usepackage{url}

\usepackage{todonotes}

\begin{document}
%
\title{Lessons Learnt in Conducting Survey Research}




%
\author{\IEEEauthorblockN{Marco Torchiano\IEEEauthorrefmark{1},
Daniel Méndez Fernández\IEEEauthorrefmark{2},
Guilherme Horta Travassos\IEEEauthorrefmark{3}, and 
Rafael Maiani de Mello\IEEEauthorrefmark{4}
\IEEEauthorblockA{\IEEEauthorrefmark{1}Politecnico di Torino, Italy\\
marco.torchiano@polito.it}
\IEEEauthorblockA{\IEEEauthorrefmark{2}Technical University of Munich, Germany\\
daniel.mendez@tum.de}
\IEEEauthorblockA{\IEEEauthorrefmark{3}COPPE/Universidade Federal do Rio de Janeiro, CNPq Researcher, Brazil\\
ght@cos.ufrj.br}
\IEEEauthorblockA{\IEEEauthorrefmark{4}Informatics Department, PUC-Rio, Brazil\\
rmaiani@inf.puc-rio.br}}
}


\maketitle

\begin{abstract}
Context: Surveys constitute an valuable tool to capture a large-scale snapshot of the state of the practice. Apparently trivial to adopt, surveys hide, however, several pitfalls that might hinder rendering the result valid and, thus, useful.

Goal: We aim at providing an overview of main pitfalls in software engineering surveys and report on practical ways to deal with them.

Method: We build on the experiences we collected in conducting many studies and distill the main lessons learnt.

Results: The eight lessons learnt we report cover different aspects of the survey process ranging from the design of initial research objectives to the design of a questionnaire.

Conclusions: Our hope is that by sharing our lessons learnt, combined with a disciplined application of the general survey theory, we contribute to improving the quality of the research results achievable by employing software engineering surveys.

\end{abstract}


%
\IEEEpeerreviewmaketitle

\section{Introduction}
Software engineering (SE) concerns the systematic development of software products and over the last decades the SE communities have developed a plethora of methods and tools used to build software. Unlike disciplines concerned more with the production of physical products, the focus in software engineering is on development yielding in outcomes that are typically unique. Since the design and development of software thus mostly follows an essentially unique process, it is difficult to derive general principles and laws for software engineering.

To reason about our discipline as well as to recognise and understand the effects of tools, methods, and approaches in this inherently complex socio-technical environment, we need different empirical research methods. These allow us to explore the state of practice and validating new tools or approaches. Some of these methods, such as experiments, enable us to investigate isolated phenomena in contexts detached from reality; others, in turn, broaden the scope while taking a wide-angle snapshot of the state of the practice. One research method to be named here is survey research. Surveys have received much attention in research and practice for many years as a tool to systematically analyse opinions, experiences, expectations among the investigated populations. They allow us to ask descriptive questions (``what is happening?'') or explanatory questions (``why is this happening?''),  mostly in an exploratory way to get an initial, broad picture about the state of the practice. Exploratory surveys have been conducted in different software engineering research topics~\cite{deMello2015}, many of them serving the purpose of gathering opinions from SE professionals. The selection of the most appropriate research method depends on the expected outcome, the purpose, and the epistemological approach~\cite{Wohlin2015}.

Surveys seem apparently trivial to design and fast to conduct, yet survey research hides various pitfalls that might hinder rendering the result valid and, thus, having conducted a survey might quickly become a waste of time or, worse, lead to false conclusions. The issues most often criticised by peers~\cite{CESI2013} are: the lack of novelty in the research questions addressed, the potential limitation of the geographic scope, and the limited representativeness of the sample.

As many misconceptions and pitfalls still dominate survey research, we still need a better understanding of various facets for applying it in our field~\cite{zannier2006success}\cite{StatusSE2007}; for instance, key challenges include ensuring the representativeness of samples including the search for adequate populations and stimulating their participation, especially among practitioners in the industry~\cite{deMello2016}, or how to address the effective participation of subjects when preparing suitable questionnaires~\cite{Linaker2015}. 

In this experience report, we discuss practical challenges and selected lessons we learnt in conducting survey research in industry. Our work emerges from the organisation of the \emph{14th International Advanced School of Empirical Software Engineering}, which the authors held in conjunction with the \emph{International Symposium on Empirical Software Engineering and Measurement (ESEM)} in 2016. Our goal is to provide a summary of the most important lessons we learnt ourselves while conducting surveys in the industry to help especially inexperienced researchers facing their first surveys in industry. 

After a brief introduction to the basic concepts of survey research, we subsequently report on our experiences made along the phases:
(i) defining research objectives and the target population,
(ii) sampling,
(iii) designing a questionnaire, and
(iv) recruiting.

Our experiences emerge from a series of different surveys we conducted in industrial settings and cover, in particular, methodological aspects of the whole process. They also come out of the research efforts for designing a conceptual framework to support the identification of representative samples in surveys in Software Engineering~\cite{deMello20162}. The main surveys we conducted and on which we base our experiences are, if not limited to:
\begin{compactitem}
\item a multi-national survey on Off-The-Shelf development practices~\cite{li2008state},
\item a national survey on software migration~\cite{Torchiano2011},
\item a national survey on Model-Driven-Development~\cite{torchiano2013},
\item an international survey on the impact factors of software requirement activities~\cite{deMello2013},
\item a multi-national family of surveys on requirements engineering practices and problems~\cite{MW14}, and
\item an international survey on characteristics of agility and agile practices in software processes~\cite{deMello20152}.
\end{compactitem}

In this report, we put emphasis on design issues, i.e. pitfalls that might appear at the beginning during the design of the survey and the recruitment. Please note that we do not claim that there are not other valuable sources on how to cope with the various challenges imposed by survey research. Related work ranges, in fact, from studies on issues related to the design of surveys, e.g. most common criticisms found in peer evaluation of survey studies~\cite{CESI2013}, to guidelines for conducting surveys in software engineering (e.g.,~\cite{Linaker2015, KitchenhamPfleeger2008}). 

Our intention is to complement this work by adding our own personal experiences in conducting survey research in industry while putting emphasis on those aspects, we found to be of particular importance in our own experience. Our hope is that by sharing our experiences, combined with a disciplined application of existing guidelines, we contribute to improving the quality of the surveys conducted in industry.

\section{Survey Research: Basic Terms and Concepts}
Among the large family of empirical research methods, surveys can be classified in the ''observational'' group. Surveys constitute a means to observe from the outside essentially, i.e. surveys do not entail an active intervention component -- differently than, e.g., an experiment where one or more factors are applied and controlled.

A possible definition of the survey method is: 
\emph{a systematic observational method to gather qualitative and/or quantitative data from (a sample of) entities to characterise information, attitudes and/or behaviours from different groups of subjects regarding an object of study}~\cite{GrovesEtAl2009}.

The survey methodology can be seen from two distinct and complementary perspectives:

\begin{itemize}
	\item[\textbf{Measurement}:] focuses on assessing specific constructs of interest,
	it consists of collecting measures (values) to characterise attributes
	of the objects of interest.
	\item[\textbf{Representation}:] concerns the goal of describing
	the characteristics of a population of interest faithfully.
\end{itemize}

\begin{figure}[htbp]
	\begin{center}
		\includegraphics[width=\linewidth]{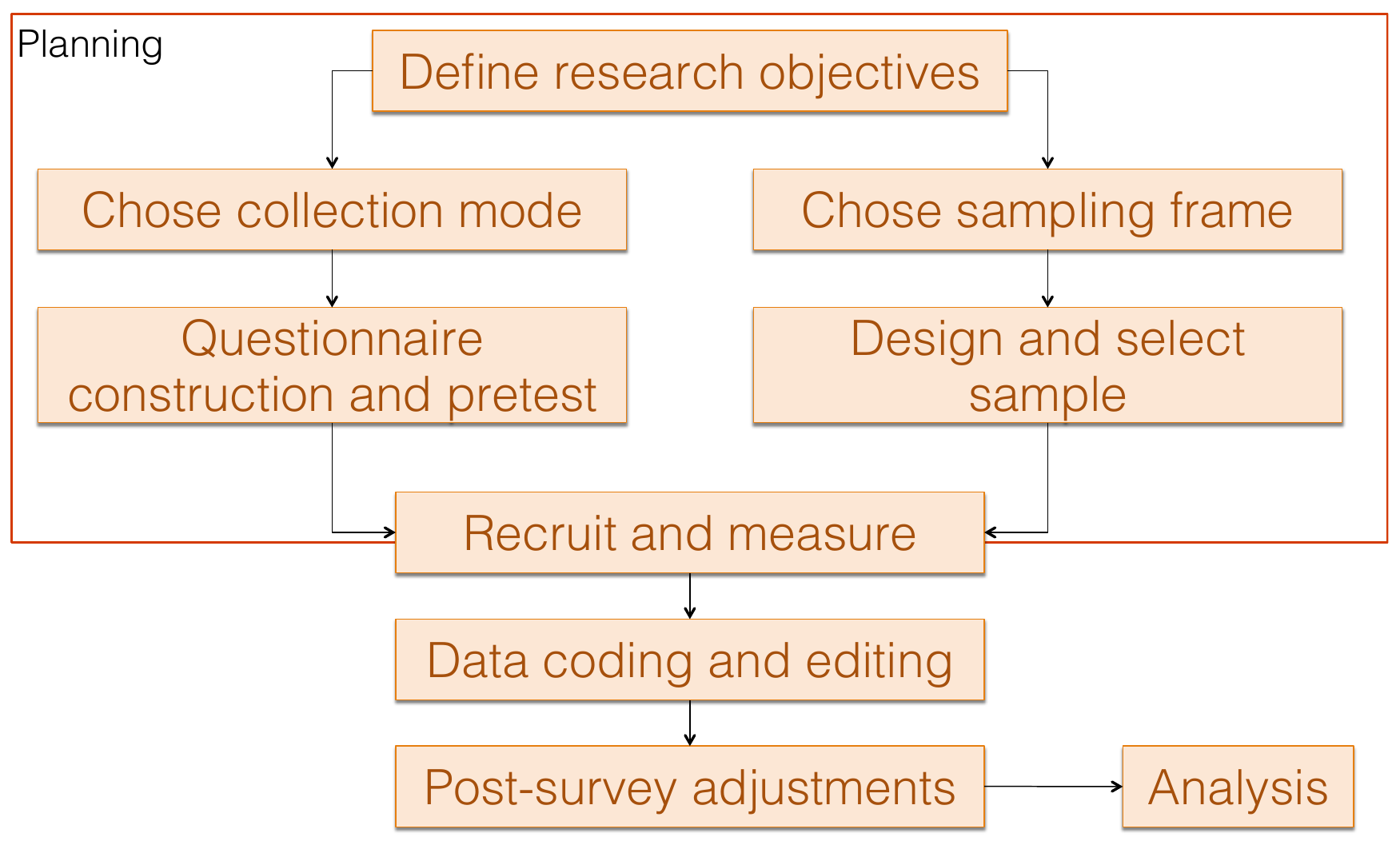}
		\caption{The survey process (adapted from~\cite{GrovesEtAl2009})}
		\label{fig:process}
	\end{center}
\end{figure}

The procedure to conduct a survey, as shown in Figure \ref{fig:process}, reflects the two above perspectives, especially in the planning phase.

The definition of the research objectives requires the definition of constructs -- the characteristics to be described -- and identification of the target population -- the subjects whose characteristics ought to be represented.

At the measurement side, a particular collection mode, e.g. interviews, paper-based questionnaire or on-line questionnaire, has to be selected, and the questionnaire must be designed and developed to measure the constructs defined earlier. At the representation side, there are two key steps: (i) the selection of a sampling frame, i.e. the instrument used to identify the members of the target population, and (ii) the design and determination of the sample, i.e. the subset of members that will be contacted.

The execution of the survey consists of recruiting the subjects and the administration of the questionnaire. Once the data collection is completed, we need to carry out the data coding and editing, and we may need conduct post-survey adjustments; for instance, to adapt the sample demographics to those of the original target population.

\section{Lessons Learnt}
In the following, we report on our lessons learnt when conducting survey research with a particular focus on surveys carried out in industrial settings. To this end, we follow the basic process illustrated in Fig.~\ref{fig:process} and report for each prominent challenges therein:

\begin{itemize}
\item description of the challenge,
\item why it matters and what the impacts are, 
\item how it can be faced (how we used to do it), and
\item a discussion on the implications and open issues.
\end{itemize}

\subsection{Defining Research Objectives}
\label{sec:research}
When it comes to planning a survey in general and setting the research objectives, in particular, we experienced a plethora of challenges of which we summarise the ones most important to us in the following. 

\subsubsection{Awareness of the Scope and Limitations of Survey Research}
Survey research plays a special role in empirical software engineering as it provides a means to distill the subjective (and often fuzzy) opinions of the respondents as opposed to reliable, hard facts or even causal relationships between phenomena. The information we opt for with survey research is often about opinions, expectations, and experiences made by the respondents in their very own environment over a long time span. 
Thus, survey research comes close to folklore gathering. While this poses several threats to the validity, it also constitutes the major strength of survey research. In contrast to, e.g., controlled experiments which need to be detached from reality using abstraction and simplification of the phenomena we investigate, surveys provide the possibility to reveal exactly this kind of subjective information we often cannot access otherwise. We experienced it to be of particular importance to understand this difference in the scope and the limitations of survey research during the planning phase when deciding for appropriate empirical methods.

\paragraph*{Lessons Learnt}

Especially novice researchers tend to have a better understanding of experiments than of, generally speaking, research methods that rely on a mix of quantitative and qualitative data. Too often, they use survey research to focus on aspects that can be quantified by numbers. Such a quantification then can lead to the illusion of treating the data as if it was revealed via experimentations, seeing causal dependencies where there might be none and over-interpreting the findings. Survey research can provide interesting insights when focusing on their strengths while having an awareness of their limitations. That is, instead of using surveys to ask normative questions ("What should we do?"), they should ask descriptive questions ("What is happening in your environment?"), and especially explanatory questions ("Why is this happening?"). Over and above all, surveys can help us getting a better understanding of what the opinion, the perception, and the experiences of the respondents are. They constitute an effective means to ask "Why"-questions which we are usually not able to investigate through experiments.

\subsubsection{Identifying the Target Population}

When designing a survey, the main question is which should be its target population. The choice of the target population is crucial. The adequacy of respondents to provide significant answers decides upon the success of a survey. 

\paragraph*{Lessons Learnt}
We experienced it to happen too often that researchers decide upon the target population based on the hopes to get high response rates -- i.e. availability and willingness of people to participate in the survey -- rather than whether they are most appropriate to provide accurate answers. If, for example, we design a survey on  problems experienced in requirements engineering, it makes little sense to distribute the survey among architects only or all possible project roles just because these are the people at our disposal. If on the other hand, we design a survey to evaluate the research outcomes of a whole community from a practitioners' perspective, it makes little sense to ask practitioners from a single company only. Thus, the choice of the target population is crucial because it decides upon the (scope of) validity of a survey.

In our experience, it is important to avoid restricting the target population based on factors such as availability. The question "Who can best provide you with the information you need?" should drive the selection rather than "Who are the subjects most likely available to participate?". It is easier to handle a small response rate than useless responses.

\subsection{Target Population}
\label{sec:target}
Once we decided upon the scope of the survey and the coarse target population, we need to characterise it. 

\subsubsection{Identifying the Unit of Analysis}

The survey unit of analysis typically is the individual, i.e. each subject part of the target population. In some cases, the unit of analysis may be a particular group of individuals, such as organisational units or project teams \cite{deMello2015}. For example, when investigating the opinions of developers, we should rely on set(s) of individual developers who can even be further refined (to, let's say, Java developers). Or, for instance, when investigating how software companies define their requirements engineering process models, we should rely on set(s) of software companies, each represented by one or more employees.

The establishment of the entity that will identify each unit of analysis should take into account the survey research objective. Once the unit of analysis is established, it will guide sampling activities. Surveying a random sample composed of developers is different from surveying developers from a random sample of software houses. The former sample may be helpful to answer general research questions, e.g. general experiences made in code debugging, while the latter  may be useful to investigate specific issues dealt with by different software houses, e.g. procedures followed for debugging code. Therefore, the establishment of the adequate unit of analysis is crucial although often underestimated. Having the wrong units of analysis may render the survey data useless for the addressed questions.

\paragraph*{Lessons Learnt}
The identification of the units of analysis should be exclusively guided by the survey research objective and the related research questions; also, we should check whether the perspective taken by the respondents suffices to give proper responses to those questions. As so often, the pitfall lies in the details and deserves proper reflection. For example, when investigating which programming practices are \emph{preferred} by particular developer groups (e.g. Java developers), we might approach Java programmers independent (to some extent) of their organisational context. We might use community-specific mailing lists and ask questions on their typical projects, but quickly focus on their preferences and possible reasons. In contrast, when investigating which programming practices are \emph{followed} by specific developer groups (e.g. Java developers), we might approach Java developers grouped by their organisational context as this context defines the software process models and the guidelines they (should) follow. We might even change the way we distribute the survey and approach various specific companies, structured according to multiple domains rather than whole communities.

The other side of the medal entails a further potential pitfall: too often we are guided by the chances to get proper responses rather than by the possibilities of asking interesting questions. One lesson we learnt includes to first think about the research questions we are interested in answering and then identifying the type of units of analysis we need, never vice versa. 

The low incidence of surveys in software engineering having groups of individuals as units of analysis may be explained by the difficulty in identifying representative samples, especially when involving organisational units. Although there are some resources available that may help to mitigate such limitation; they are typically driven by individuals apart from organisations or working teams \cite{deMello2015}. Thus, such limitation should be taken into account early on when designing a survey.

\subsubsection{Characterising the Units of Analysis and Subjects}

Once the units of analysis are identified, they need to be properly characterised. Different research objectives may demand different attributes to characterising individuals or groups of individuals in the survey. Such attributes include demographics but typically go further~\cite{deMello2015}. A representative sample might be identified by stratifying the target population w.r.t. its descriptive attributes and randomly sampling from the resulting strata~\cite{deMello2016}\cite{Conradi2005}. 

Thus, the use of the correct attributes will support the following sampling activities and also the generalisation of the results.  

\paragraph*{Lessons Learnt}
Different units of analysis can be characterised by different means. For example:
\begin{compactitem}
    \item \emph{Individuals} can be characterised through attributes such as experience in a particular research context, experience in software engineering field or particular role, current professional role at the very moment of running the survey, location and higher academic degree or cultural background.
    \item \emph{Organisations} can be characterised via attributes such as size (scale typically based on the number of employees), industry segment (software factory, avionics, finance, health, telecommunications, etc.), location or organisation type (government, private companies, universities, etc.).
    \item \emph{Project teams} can be characterised via attributes such as team size; team distribution along different regions and time zones, client/product domain (avionics, finance, health, telecommunications, etc.) or the family of systems envisioned (embedded reactive system or business information systems).
\end{compactitem}

Generalisation of results is further a challenging task in empirical research, especially in survey research, and the ability to reason is very closely related to the characteristics we capture about the units of analysis. We recommend relying on standards where possible as they can be particularly useful to provide scales and even nominal values. For instance, CMMI-DEV maturity levels can be used to characterise organisational units regarding their maturity in the software process. The roles as defined by the Rational Unified Process can be used to characterise subjects’ current positions and the typical responsibilities they have. Software projects can be described by the variables used in the (software process) tailoring mechanisms of the methodologies. Further, the use of standards supports the comparison with other studies and the corroboration of our results with existing evidence, respectively synthesising results obtained from multiple surveys.

\subsection{Sampling}
\label{sec:sampling}
Once the target population and the unit of analysis are characterised, the survey sampling frame should be established, and the sampling design should be defined.

\paragraph*{Lessons Learnt}

Suitable sampling frames are rarely available in SE research. It is, therefore, important to carefully search for candidates of population sources. We should avoid the convenience in searching candidates, trying to answer: \emph{Where is a representative population from the survey target population available?} However, the term \emph{representative} is crucial here as it is uncommon to have census data available regarding software engineering professionals and organisations. It is, therefore, important to have in mind that this question cannot be answered based on externally available references only, but that we have to rely on otherwise relevant attributes already defined to characterise the units (see Sect.~\ref{sec:target}). For instance, although we do not have census data about the requirements engineers working in Germany, we still may assume that experience in requirements engineering, academic degree, and domain could be essential attributes to conclude that a frame population is potentially representative. Thus, it is important to make such assumptions explicit when reporting the results.

The set of population sources available to support software engineering research is, in fact, a universe of imperfect alternatives. They can rely, for example, on attendees of practitioner conferences, social networks, discussion groups, yellow pages, or organisation members. Most of these sources have not be designed to support software engineering research. For instance, projects repositories and worldwide professional social networks can be helpful to identify representative populations composed by SE professionals. However, it is important to have in mind that the policies to access the content available may change at any time. On the other hand, catalogues provided by recognised institutes, associations, and governments to retrieve relevant sets of SE professionals may be useful and its access more stable. For instance, the SEI (\url{www.sei.cmu.edu}) provides an open list of organisations and organisational units certified in each CMMI-DEV level. 

We experienced the following essential requirements for a good source of population (see also~\cite{deMello20162}). The source should
\begin{compactitem}
	\item not intentionally represent a segregated subset from the target population, 
	\item not present any bias on including on its database preferentially only subsets of the target population. Unequal criteria for including search units mean unequal sampling opportunities,
	\item allow for identifying all units by a distinct logical or numerical id,
	\item allow for accessing all units (if there are hidden elements, it is not possible to adequately contextualise the population).
\end{compactitem}

After identifying a suitable population source, we need to establish procedures to depict the survey sampling frame systematically. Such a practice is necessary to support understanding the generalisability of the results and also support future survey replications. Apart from the sampling design applied, the sample size is a particular challenge in software engineering. Our experience shows that in the case of voluntary surveys having professionals from industry as subjects, it is important to establish significantly higher sample sizes, considering the expectation of a very low participation rate (typically less than 10\%).

\subsection{Questionnaire}
\label{sec:questionnaire}
The questionnaire is the instrument that is used to measure the constructs that are investigated through the survey. The aim is to design a clear, simple and consistent set of items to compose the questionnaire.

It is important to invest effort in the design because bad questionnaires can lead subjects initially willing to participate to give up, besides the possibility of confounding the results. The questions can assume different structures for its items: closed ones (a set of predefined answers), open ones (free answer), and semi-closed ones (combining a set of predefined answers with an open answer). At least, four quality perspectives should be applied when designing the survey questionnaire: \emph{simplicity, clarity, effortless,} and \emph{responsiveness}. The challenge is to assure their balance when designing the questions~\cite{Linaker2015}.

\subsubsection{Simplicity}

The questions should use simple and appropriate wording for the survey
questions. It is important to reduce both understanding effort and (syntactical) ambiguity 

\paragraph*{Lessons Learnt}
It can be reached by avoiding technical terms as much as
possible or defining them in the questionnaire, according to the survey
target population. As much as possible, prefer to design short questions
regarding a single concept and avoid double-barrelled questions, which
can confound the answers. 
Vague sentences should also be avoided while writing survey questions.

Passive voices and double negative questions should be avoided because they naturally require more time to be understood and increase the risk of incorrect answers.
In general, all available recommendations for writing natural language specifications should be followed.

A counter-example of statement agreement question where two distinct constructs are merged in a single question is:
"\emph{In your opinion}, do you \emph{agree or disagree} that \emph{code
refactoring is a need as well as code smell detection?}".
A better alternative would be "\emph{Code refactoring is an essential practice 
for improving the understanding of object-oriented code.}"
A double negative question is, for instance, "Which of the following NFR \emph{do you disagree are not relevant} in
the context of real-time systems?"

\subsubsection{Clarity}

The question should be (semantically) unambiguous and allow for concrete answers.
In particular, it should refer to facts the respondents do know.

\paragraph*{Lessons Learnt}
The survey designers should avoid biased questions. An example of a biased item is: "Do you prefer working in projects following agile methods or those following \emph{usual non-agile} approaches?".
Avoiding bias can be achieved by only phrasing questions that do not suggest likely answers or
responses. Besides biased questions, sensitive questions should be avoided in the 
questionnaire as they might take respondents out of their comfort zone so that they will probably opt out of the survey. 
Such sensitive questions can include in which company the respondents work, what respondents' income are, or opinions about their organisation or their management. Also, asking \emph{age, gender, and marital status} for characterising requirements engineers clearly addresses sensitive aspects; if useful to answer the particular research questions, the relation should be made clear to the respondents.

\subsubsection{Effortless}

Industry participants usually tend to participate more in surveys when they see the value of the survey for their daily work, yet a survey should still not demand too much effort as answering surveys is something the respondents do besides their daily work. Demanding questions increase the time to react and affect the willingness of respondents to complete the survey.

\paragraph*{Lessons Learnt} 

One important aspect to consider is that the questions should deal with recent events rather than events that lie far in the past so to the questions allow respondents to give quickly clear answers.

Moreover, respondents should not be asked to provide too many specific details.

A question that considers events in the far past is, for instance, "Considering \emph{the main characteristics of the last ten software projects you have worked on}, please answer the following questions:\ldots{}"

An example for a demanding question is "After \emph{reading the attached papers regarding non-functional requirements} (NFR), please answer the following questions:\ldots{}".

\subsubsection{Responsiveness}
The response format is not only necessary from the perspective of the respondents answering the survey questions, but it is also of particular importance for the data analysis. Answers should, therefore, be given by using an adequate response format. Items with inappropriate response formats may hinder applying relevant statistical tests, significantly (and unnecessarily) increase the efforts regarding data analysis or even render the answers impossible to analyse and interpret in a meaningful manner. A summary of the most common item response formats
is reported in Table \ref{tab:responsiveness} with their main
features. 

\paragraph*{Lessons Learnt} 

Whenever an ordinal scale is used, it is important to label exactly all the options. For instance, an item that asks to rate the level of confidence on a scale from 1 to 6 makes it difficult to answer; e.g., assuming that 4 to 6 are high levels, what is precisely the difference between 4 and 5? Such problem can lead to inaccurate measures.

We also learnt that it is useful to use Likert scales when a construct can be represented as a general agreement with a statement. These come, however, as a measure with limited precision.

It is further important to express explicit references to allow an easy judgment by the respondent. For instance, to assess the experience with a programming language, the least precise option consists in using a Likert item response set for the statement "\emph{I have high experience in using the X}". A better question would be "\emph{What is your experience with X?}" with an ordinal response set ranging from "none" to "very high". In our experience, the best option is to use ranges as shown in Table~\ref{tab:responsiveness} where clear reference points are defined.

\begin{table*}[tb]
\centering
	\caption{Main type of response scales}
	\label{tab:responsiveness}
	\begin{tabular}{@{}lp{0.3\linewidth}p{0.5\linewidth}@{}}
		\toprule
		\textbf{Response Format} & \textbf{Features} & \textbf{Example}\tabularnewline
		\midrule
		Nominal & Closed questions
		
		Statistical analysis based on frequency & Do you have experience in Java
		programming?
		
		( ) Yes ( ) No
		\tabularnewline\addlinespace
		Ordinal / Likert Scale & Closed questions
		
		Not necessarily equally distributed intervals
		
		Significantly restricts statistical analysis & How much experience do
		you have in Java programming?
		
		\begin{enumerate}
			\def\labelenumi{\alph{enumi})}
			\item
			Very High experience
			\item
			High Experience
			\item
			Little Experience
			\item
			Very Little experience
		\end{enumerate}\vspace{-0.8em}\tabularnewline
		Ranges & Closed questions
		
		Ranges are considered equally distributed
		
		Statistical analysis is less restrictive than Ordinal Scale & How much
		experience do you have in Java Programming?
		
		\begin{enumerate}
			\def\labelenumi{\alph{enumi})}
			\item
			Less than one year
			\item
			1 year to 3 years
			\item
			3 years to 5 years
			\item
			More than 5 years
		\end{enumerate}\vspace{-0.8em}\tabularnewline
		Free-Text & Open questions
		
		Qualitative content analysis (e.g. via manual coding)
		
		High effort on data analysis & How much experience do you have in Java
		programming?
		
		\emph{I have been working with Java programming at companies since
			2011. Before, I got my first Java certification in 2009, when I started
			working on personal projects. However, I have difficult with
			object-oriented parts\ldots{}}
		\tabularnewline\addlinespace
		Numeric values & Open/Closed questions
		
		Allow a wide range of statistical analysis & How much experience do you
		have in Java programming?
		
		\_\_5\_\_ years\tabularnewline
		\bottomrule
	\end{tabular}
\end{table*}

\subsection{Recruiting}
\label{sec:recruit}

Once the target population has been defined and sampled, and the instrument has been designed, the invitation message has to be delivered to the right person.  It includes questions ranging from whether to target known individuals in closed surveys (e.g. project leads from existing contacts) to whether to target a broad audience in open surveys where the distribution of the invitation could be supported in public forums, mailing lists, and even social media. All these questions pose challenges that need to be addressed already in the design (e.g. in demographical questions in case of open surveys). However, two aspects that turned out to be relevant in our experience are the control of who is recruited and the incentives to let people respond to the questionnaire.

\subsubsection{Controlling participation}

When the sample is defined, the recruitment must comply with the sample characteristics. The usual way to invite the members of the sample is by sending individual or general messages. In any case, it is important to restrict the survey access to the individuals originally recruited. The characteristics of an uncontrolled population could significantly differ from the intend one.

\paragraph*{Lessons Learnt}
We consider it to be imperative to take into account the local policies and laws, the possible rewards, the collected personal information, and the ethical issues which altogether must conform to institutional or national rules. This aspect must be specifically cared for and explicitly conveyed to the participants. 

Further, we experienced it to be of particular importance to use a direct communication and avoid what can be called a ``spreading spree'', e.g., through mailing lists, forum invitation messages, or crowdsourcing tools (e.g. Amazon MechanicalTurk). Although they offer the advantage to increase the visibility of the survey, they still pose the problem that it is impossible to control who will read the invitation message, thus, hindering the control of the recruited sample. For instance, in one of our surveys~\cite{torchiano2013}, we have sent out the invitation (in addition to framing the population using a commercial firm's database) to a mailing list of attendees of a large conference. In such a case, it was not possible to compute any response rates and an extra effort was required to compare the features of the responding population to that of the target population. 

Finally, especially in closed surveys forwarding the invitation message should either not be allowed or only in a controlled manner (e.g. by providing clear instructions for doing so). Forwarding invitations could extend the sample in an unforeseen way. One way to deal with this problem is via technical mechanisms as most web-based questionnaire tools allow the definition of unique tokens to the population.

\subsubsection{Stimulating participation}

A critical issue in the recruitment phase consists of stimulating the sampled participants to respond to the questionnaire.
Participants stimulation is of particular importance to achieve high response rates. The initial invitation to participate in the survey should, therefore, be sent to the sampled participants and provide a general description that encourages responding. Often, however, the participants are not able to complete the questionnaire at the very moment they receive the invitation message. So it is likely they delay the time to fill in the questionnaire to the next moment so, it is possible they forget about it later on.

Therefore, stimulation has not only the goal of increasing the chances of responses after sending out the invitation, but also along the whole data collection phase.

\paragraph*{Lessons Learnt}
Although we are far from really understanding how to motivate respondents to participate in a survey, we experienced different factors as useful means to boost the participation \cite{deMello20162}. It includes the initial recruitment invitation as well as potential reminder messages sent out in a later stage.

Concerning the \emph{initial invitation}, it is important to include in the invitation message an observation 
	explaining the relevance of subject participation, to encouraged participation. The general context of the study should be briefly explained. For instance, if investigating a particular tool or technique, it is interesting to know how many respondents do actually use it (the adoption rate). However, if a tool is already
	mentioned in the invitation, then people not using it tend to self-exclude
	themselves: it brings to an under-estimation of the adoption rate.
Besides, it is better to establish a limited and not too long period of time to answer the survey, e.g. a few weeks only; if postponing, a closer date makes forgetting less likely.
Moreover, it is possible to offer rewards (e.g. raffles, payments, sharing results); this makes
	the less motivated participants more likely to be involved, thus avoiding
	a possible self-exclusion bias. However, we experienced the biggest reward for practitioners to see the value in the survey for their own work and, thus, we encourage researchers to share the results of the survey with the practitioners afterwards.

Concerning the \emph{reminders}, 
they should be used with care; a potential participant that is willing to participate might be upset by, e.g., an impolite reminder message. Avoid sending too many reminders; while a first reminder (typically between 4 and 8 days) may significantly increase the response rate, a third reminder can improve the response rate by a small amount only.
Avoid reminding those who already have participated: a reminder message sent to all participants is ineffective and may annoy those who already responded making them less likely to respond to future surveys.

\section{Conclusions}
Survey research has become an indispensable tool in software engineering research as it allows us to explore the state of practice taking a broader perspective. Although survey research is wide-spread nowadays especially to distill opinions, experiences, and expectations among practitioners, it still poses a plethora of more practical challenges which are not typically addressed in available guidelines.

In this paper, we shared our personal lessons learnt in conducting survey research in industry. These lessons emerged from a series of surveys and investigating methodological issues on planning surveys in Software Engineering. Our hope is to support especially more inexperienced researchers confronted with their first surveys by complementing available methodological guidelines with a more practical view. Of course, the lessons shared in the paper at hands reflect our very personal views and other researchers might come up with different challenges and lessons learnt. We thus encourage other researchers also to share their experiences and lessons learnt to keep strengthening our expertise as a community.

\bibliographystyle{IEEEtran}
\bibliography{biblio}

\begin{thebibliography}{10}
\providecommand{\url}[1]{#1}
\csname url@samestyle\endcsname
\providecommand{\newblock}{\relax}
\providecommand{\bibinfo}[2]{#2}
\providecommand{\BIBentrySTDinterwordspacing}{\spaceskip=0pt\relax}
\providecommand{\BIBentryALTinterwordstretchfactor}{4}
\providecommand{\BIBentryALTinterwordspacing}{\spaceskip=\fontdimen2\font plus
\BIBentryALTinterwordstretchfactor\fontdimen3\font minus
  \fontdimen4\font\relax}
\providecommand{\BIBforeignlanguage}[2]{{%
\expandafter\ifx\csname l@#1\endcsname\relax
\typeout{** WARNING: IEEEtran.bst: No hyphenation pattern has been}%
\typeout{** loaded for the language `#1'. Using the pattern for}%
\typeout{** the default language instead.}%
\else
\language=\csname l@#1\endcsname
\fi
#2}}
\providecommand{\BIBdecl}{\relax}
\BIBdecl

\bibitem{deMello2015}
R.~M. de~Mello and G.~H. Travassos, ``Characterizing sampling frames in
  software engineering surveys,'' in \emph{Proceedings of the 18th
  Iberoamerican Conference on Sotware Engineering}, ser. CIbSE 2015, 2015.

\bibitem{Wohlin2015}
C.~Wohlin and A.~Aurum, ``Towards a decision-making structure for selecting a
  research design in empirical software engineering,'' \emph{Empirical Software
  Engineering}, vol.~20, no.~6, pp. 1427--1455, 2015.

\bibitem{CESI2013}
M.~Torchiano and F.~Ricca, ``Six reasons for rejecting an industrial survey
  paper,'' in \emph{Conducting Empirical Studies in Industry (CESI), 2013 1st
  International Workshop on}.\hskip 1em plus 0.5em minus 0.4em\relax IEEE,
  2013, pp. 21--26.

\bibitem{zannier2006success}
C.~Zannier, G.~Melnik, and F.~Maurer, ``On the success of empirical studies in
  the international conference on software engineering,'' in \emph{Proceedings
  of the 28th international conference on Software engineering}.\hskip 1em plus
  0.5em minus 0.4em\relax ACM, 2006, pp. 341--350.

\bibitem{StatusSE2007}
A.~H{\"o}fer and W.~F. Tichy, ``Status of empirical research in software
  engineering,'' in \emph{Empirical Software Engineering Issues. Critical
  Assessment and Future Directions}, V.~R. Basili, D.~Rombach, K.~Schneider,
  B.~Kitchenham, D.~Pfahl, and R.~W. Selby, Eds., 2007, pp. 10--19.

\bibitem{deMello2016}
R.~M. de~Mello and G.~H. Travassos, ``Surveys in software engineering:
  Identifying representative samples,'' in \emph{Proc. 10th ACM/IEEE
  International Symposium on Empirical Software Engineering and
  Measurement}.\hskip 1em plus 0.5em minus 0.4em\relax New York, NY, USA: ACM,
  2016, pp. 55:1--55:6.

\bibitem{Linaker2015}
J.~Lin{\aa}ker, S.~M. Sulaman, R.~Maiani~de Mello, and M.~H{\"o}st,
  ``\BIBforeignlanguage{eng}{Guidelines for conducting surveys in software
  engineering},'' Tech. Rep., 2015.

\bibitem{deMello20162}
\BIBentryALTinterwordspacing
R.~Maiani~de Mello, ``\BIBforeignlanguage{eng}{Conceptual framework for
  supporting the identification of representative samples for surveys in
  software engineering},'' 2016. [Online]. Available:
  \url{http://www.cos.ufrj.br/uploadfile/publicacao/2611.pdf}
\BIBentrySTDinterwordspacing

\bibitem{li2008state}
J.~Li, O.~Slyngstad, M.~Torchiano, M.~Morisio, and C.~Bunse, ``A
  state-of-the-practice survey of risk management in development with
  off-the-shelf software components,'' \emph{Software Engineering, IEEE
  Transactions on}, vol.~34, no.~2, pp. 271--286, 2008.

\bibitem{Torchiano2011}
M.~Torchiano, M.~{Di Penta}, F.~Ricca, A.~{De Lucia}, and F.~Lanubile,
  ``Migration of information systems in the italian industry: A state of the
  practice survey,'' \emph{Information and Software Technology}, vol.~53, pp.
  71--86, January 2011.

\bibitem{torchiano2013}
M.~Torchiano, F.~C.~A. Tomassetti, F.~Ricca, A.~Tiso, and G.~Reggio,
  ``Relevance, benefits, and problems of software modelling and model driven
  techniques{--}a survey in the italian industry,'' \emph{Journal of Systems
  and Software}, vol.~86, no.~8, pp. 2110--2126, 2013.

\bibitem{deMello2013}
R.~M. de~Mello and G.~H. Travassos, ``Would sociable software engineers observe
  better?'' in \emph{7th ACM/IEEE International Symposium on Empirical Software
  Engineering and Measurement}, Oct 2013, pp. 279--282.

\bibitem{MW14}
{M{\'e}ndez Fern{\'a}ndez, D. and Wagner, S.}, ``{Naming the Pain in
  Requirments Enginering: A Design for a global Family of Surveys and First
  Results from Germany},'' \emph{{Information and Software Technology}}, 2014.

\bibitem{deMello20152}
R.~M. de~Mello, P.~C. da~Silva, and G.~H. Travassos, ``Investigating
  probabilistic sampling approaches for large-scale surveys in software
  engineering,'' \emph{Journal of Software Engineering Research and
  Development}, vol.~3, no.~1, p.~8, 2015.

\bibitem{KitchenhamPfleeger2008}
\BIBentryALTinterwordspacing
B.~Kitchenham and S.~Pfleeger, ``Personal opinion surveys,'' in \emph{Guide to
  Advanced Empirical Software Engineering}, F.~Shull and Singer, Eds.\hskip 1em
  plus 0.5em minus 0.4em\relax Springer London, 2008, pp. 63--92. [Online].
  Available: \url{http://www.springerlink.com/content/g7m7x79l854v2736}
\BIBentrySTDinterwordspacing

\bibitem{GrovesEtAl2009}
R.~M. Groves, F.~J.~J. Fowler, M.~P. Couper, J.~M. Lepkowski, E.~Singer, and
  R.~Tourangeau, \emph{Survey Methodology}.\hskip 1em plus 0.5em minus
  0.4em\relax John Wiley and Sons, 2009.

\bibitem{Conradi2005}
R.~Conradi, J.~Li, O.~Slyngstad, V.~Kampenes, C.~Bunse, M.~Morisio, and
  M.~Torchiano, ``Reflections on conducting an international survey of software
  engineering,'' in \emph{Empirical Software Engineering, 2005. 2005
  International Symposium on}, 2005, pp. 1--10.

\end{thebibliography}

\end{document}